\documentclass[12pt,preprint]{aastex}

\def\Msun{$\rm M_\odot~$}

\def\Porb{$P_{\rm orb}$}
\def\Mwd{$\rm M_{WD}$}

\def\simgt{\lower.5ex\hbox{$\; \buildrel > \over \sim \;$}}
\def\simlt{\lower.5ex\hbox{$\; \buildrel < \over \sim \;$}}
\def\ltsima{$\; \buildrel < \over \sim \;$}
\def\gtsima{$\; \buildrel > \over \sim \;$} 
\def\lsim{\lower.5ex\hbox{\ltsima}} 
\def\gsim{\lower.5ex\hbox{\gtsima}} 
\def\msun{${\rm M_\odot}$} 

\begin{document} 
 
\title{Radio--ejection and bump--related orbital period gap 
of millisecond binary pulsars}
 
\author{F. D'Antona,
\altaffilmark{1} P. Ventura\altaffilmark{1}, L. Burderi\altaffilmark{2},
T. Di Salvo\altaffilmark{3},
G. Lavagetto\altaffilmark{3}, A. Possenti\altaffilmark{4} 
\& A. Teodorescu\altaffilmark{5} }
 
\affil{\altaffilmark{1}INAF Osservatorio Astronomico  di Roma, via Frascati
33, 00040 MontePorzio, Italy; dantona@oa-roma.inaf.it}
\affil{\altaffilmark{2}Universit\'a di Cagliari, Cagliari, Italy}
\affil{\altaffilmark{3}Universit\'a di Palermo, Palermo, Italy}
\affil{\altaffilmark{4}INAF Osservatorio di Cagliari, Cagliari, Italy}
\affil{\altaffilmark{5}Universit\'a di Tor Vergata, Roma}

\begin{abstract}  

The monotonic increase of the radius of low mass stars during their ascent on 
the red giant branch halts when they suffer a temporary contraction. This 
occurs when the hydrogen burning shell reaches the discontinuity in hydrogen 
content left from the maximum increase in the convective extension, at the time 
of the first dredge up, and produces a well known ``bump" in the luminosity 
function of the red giants of globular clusters. If the giant is the mass 
losing component in a binary in which mass transfer occurs on the nuclear 
evolution time scale, this event produces a temporary stop in the mass 
transfer, which we will name ``bump related" detachment. If the accreting 
companion is a neutron star, in which the previous mass transfer has spun up 
the pulsar down to millisecond periods, the subsequent mass transfer phase may 
be altered by the presence of the energetic pulsar. In fact, the onset of a 
radio--ejection phase produces loss of mass and angular momentum from the 
sytem.  We show that this sequence of events may be at the basis of the 
shortage of systems with periods between $\sim 20$\ and $\sim 60$ days in the 
distribution of binaries containing millisecond pulsars. We predict that 
systems which  can be discovered at periods into the gap should have 
preferentially either magnetic moments smaller than $\sim 2 \times 
10^{26}$Gcm$^3$, or larger than $\sim 4 \times 10^{26}$Gcm$^3$. We further show 
that this period gap should not be present in population II. \end{abstract}


\section{Introduction}
\label{sec:intro}
The basic evolution of low mass X--ray binaries above the `bifurcation
period' \citep{tutukov, ergma} has been often described
\citep[e.g.][]{webbink1983}.  The endpoint of this evolution is a wide
system (orbital period \Porb\ from a few to hundreds of days)
containing a millisecond pulsar (MSP, the neutron star primary of
initial mass M$_{1}$ spun up by mass transfer from the secondary, of
initial mass M$_2$) and a low mass helium white dwarf, remnant helium
core of the mass losing red giant.  The simple relation connecting the
giant stellar radius with the helium core mass of the star, which has
a very slight dependence on the stellar mass and on the chemical
composition, coupled with the Kepler law, produces a relation between
the final white dwarf mass \Mwd\ and the final \Porb, which has been
often studied \citep[e.g.][]{tauris1996, rappaport1995,
tauris-savon1999} and compared with the \Porb\ vs. \Mwd\ distribution
of the more than 60 known MSPs.  The sample of binary MSP known up
today shows a shortage of systems with orbital periods between 22 and
56 days. There have been two main attempts to explain this `period
gap'.  According to \cite{tauris1996}, the gap is related to the
bifurcation period, which he assumes to be around 2 days. Systems of
secondary mass M$_2 \simlt 1.4$M$_\odot$, and beginning mass transfer
just above 2 days period, should evolve to final \Porb$\simgt$60
days. If the secondary has an initial mass $\simgt$1.4\msun, the
binary might evolve through a second common envelope phase, ending at
P$\simlt$20 days.  \cite{taam2000} revise this picture attributing the
periods shorter than 20 days to early massive case B evolution. They
question the idea that the bifurcation period is as large as 2 days
(to allow a minimum final period larger than 60 days), as evolution
towards shorter periods for systems below (but close to) 2 days
requires a very strong magnetic braking.  \cite{taam2000} then
attribute the period gap to the small range of initial core masses
involved (0.17 $\simlt M_c/M_\odot \simlt 0.2$).  In this latter
scenario, a carbon oxygen white dwarf of mass larger than 0.35\msun\
should be present at periods shorter than 20 days.  Notice that the
two best determined masses are $0.258^{+0.028}_{-0.016}$, at
\Porb=12.3d (Kaspi et al. 1994) and $0.237^{+0.013}_{-0.022}$\msun, at
\Porb=5.741d \citep{vanstraten}, so the standard case B evolution
produces at least some systems below 20d.

We reconsider the evolution of the systems containing a neutron star
and a giant of low initial mass. We only consider the systems in which
the initial secondary mass is smaller than the primary mass, in the
range $M_{1,in}=1.2-1.35$\msun, and assume population I composition
(metal mass fraction Z=0.02)\footnote{For larger masses we have an
initial phase of mass transfer on the thermal timescale, producing an
initial shrinkage of the orbit, until the mass ratio is reversed and
the evolution proceeds to longer periods.}.

For a wide range of initial periods, the systems suffer a phase of
detachment, when the hydrogen burning shell reaches the point of
deepest penetration of the convective envelope at the basis of the red
giant branch \citep{thomas, iben}. When the hydrogen content suddenly
increases, there is a thermal readjustment of the shell physical
conditions, with a decrease in the shell temperatures and,
consequently, in the nuclear burning luminosity. Thus the star
contracts in order to adjust to the level of nuclear reactions in the
shell.  This phase is an important evolutionary tool in the single
star evolution, as the radius contraction is associated to a drop in
the luminosity along the red giant branch, producing a ``bump" in the
luminosity function of simple stellar populations, which has been
observed in many globular clusters \citep{ferraro, zoccali}. It has
also been often identified in binary evolution \citep{kipp1967,
webbink1983, kolb-ritter1990}, although never discussed in the context
of the evolution leading to MSP.

We show that the bump related detachment of the system occurs for a
good fraction of the evolutionary paths leading to long period binary
MSP, and consider the possible consequences for the orbital period
evolution.  Although we do not follow explicitly the spin--up evolution
of the accreting neutron star, we consider that,
when the system resumes mass transfer, the NS is now a millisecond
pulsar. Therefore, mass accretion on the NS may be inhibited by the
radio--ejection due to the pressure exerted by the MSP radiation on
the matter at the inner lagrangian point \citep{rst89, st90,
burderi2001, burderi2002}. This alters the binary evolution, as the
mass is lost from the system carrying away a specific angular momentum
(AM) larger than the average. In fact, the AM lost should be of the
order of magnitude of the AM of the donor, whose mass at this stage is
only 0.4--0.5\Msun, much smaller than the NS mass.

We model parametrically the evolution in the phase of radioejection,
and show that it naturally leads, in many cases, the final periods of
systems away from the range 20--60 days.  We also show that, if the
onset of radio--ejection is the mechanism which produces the period
gap, this gap should not be found in population II.  Unfortunately,
the orbital period distribution in globular cluster seems to be
altered by destruction mechanisms at long orbital periods, so that
this hypothesis by now can not be tested.  As a byproduct, we also
show that the relation \Porb\ vs. \Mwd\ obtained for population I in
the framework of the computed models is consistent with the three best
observational data.

\section{The standard evolution for Population I}
We build up our stellar models by the ATON2.1 code, whose input
physics is described in \cite{ventura1998a}, while the binary
evolution routines follow the description in \cite{dantona1989}. The
code has been recently updated also including explicitly the evolution
of the NS \citep{lavag2004}.  Models of 1.1, 1.2 and 1.3\msun, with
helium mass fraction Y=0.28 and metallicity Z=0.02, were evolved as
donors in binaries containing a NS companion of initial mass in the
range 1.2--1.35\msun. The initial separation was chosen to allow mass
transfer at different stages along the evolution. 
We do not discuss the previous evolution of these putative systems, that is the formation
of the neutron star via the supernova (SN) ejection, and whether it is possible to obtain
a neutron star in circular orbit with a low mass companion at the periods from
which we begin considering the mass transfer. In particular, the common envelope
phase preceding the SN event is likely to end with short orbital periods
for low mass companions, and long
period systems (e.g. P$>$100d) would be created mainly as a consequence of strong natal kicks
\citep[see, e.g][]{podsi2004} and have strong eccentricities.

In addition to
nuclear evolution, we assume that magnetic braking is active and
describe it according to \cite{verbunt1981}, with the {\it f}\
parameter fixed at 0.5. We also computed several evolutions of systems
of M=0.9\msun\ and 1.1\msun, for Y=0.24 and Z=10$^{-3}$\ to test the
different behavior of population II systems. Table 1 reports some 
parameters for the binary evolutions starting from different initial 
models along the evolution of the 1.1\msun\ as single star, 
labeled by the model number N$_{in}$. 
Smaller N$_{in}$ thus correspond to earlier 
evolutionary stages. We computed ``standard"
evolutions, which, at least during the first phase of mass transfer,
neglect the possible role of the pulsar. A series of models (labeled
by ``con" in Table 1) assume conservative mass transfer below the
Eddington limit. The matter exceeding the Eddington limit was assumed
to be lost from the system, carrying away the specific AM (j$_1$) of
the NS.  We considered also evolutions, (labeled by ``eta") in which
only a fraction $\eta=0.5$\ of the mass lost by the donor was
transferred to the NS, and the remaining half was assumed to leave the
system, carrying away the specific AM j$_1$.  We add several
evolutions in which we start again from the bump--related detachment,
and assume that all the mass lost from the donor is lost from the
system, taking away a large fraction (from 60\% to 100\%) of the
specific AM j$_2$ of the donor: this is the way in which we simulate an
efficient onset of radio-- ejection, as this mechanism stops the
matter at the inner lagrangian point.
Initial and final masses of the components and 
final periods are also reported in Table 1. 

\section{The white dwarf and neutron star masses and the final orbital periods}
\begin{figure}
\includegraphics[width=11cm]{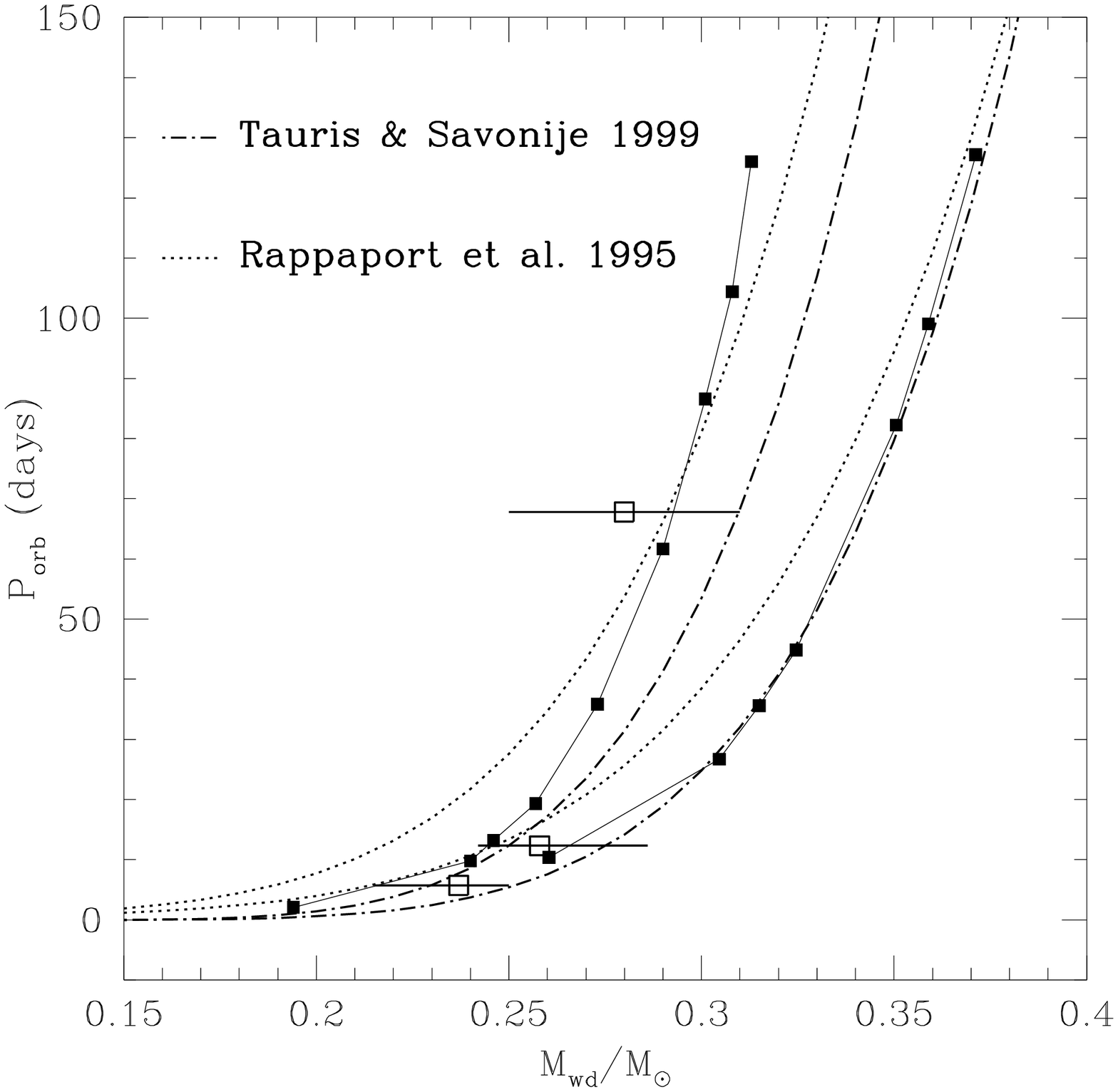}
\figcaption[f1.eps]{
In the plane \Porb\ vs. \Mwd, the result of our
computations (full squares) are compared to the analytic relations by
\cite{rappaport1995} and \cite{tauris-savon1999}.  The three curves on the
left represent the results for population I, and the three curves on
the right are for population II.  The three open squares with error
bars show three MSP binaries for which masses of the white dwarf
component are well determined: PSR J1713+0747 (Splaver et al. 2005) at
P=67.825d, B1855+09 (Kaspi et al. 1994) at P=12.327d, and J0437-4715 (
van Straten et al. 2001) at P=5.741d. These three systems are well
compatible with the white dwarf mass vs. period relation of our models.
\label{fig1}}
\end{figure}

\begin{figure}
\includegraphics[width=12cm]{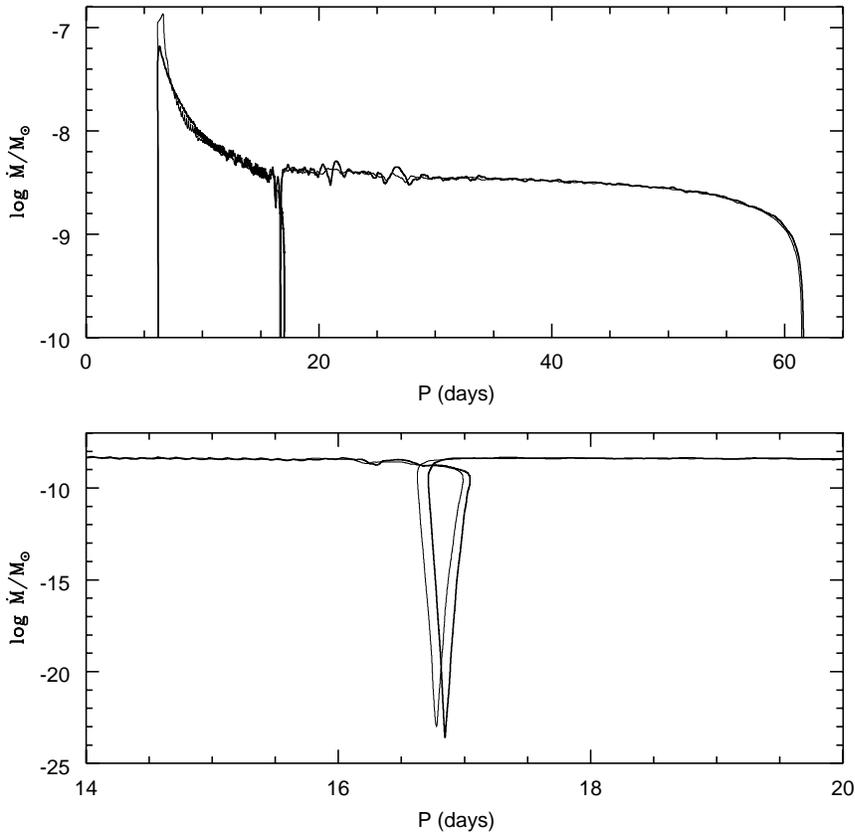}
\figcaption[f2.eps]{Comparison of  mass transfer rate vs. period for
evolutions 5con and 5eta, differing for the modalities of mass
transfer.  The bottom panel shows an enlargement at the detachment
period. During detachment the orbital period decreases due to the
magnetic braking.
\label{fig2}}
\end{figure}

Fig.\ref{fig1} shows the final \Porb\
vs. \Mwd\ relation for the population I evolutions of Table 1 (full squares on
the left) and for the additional population II models (full squares on the
right), and compares them with the analytic approximations provided by
\cite{rappaport1995} and \cite{tauris-savon1999} for population I and
II. Our data compare well with the population II relation by
\cite{tauris-savon1999}, but it is steeper than theirs at periods
$\simgt 20$days for population I.  We notice that our population I
results are in good agreement with the three observed data points
displayed in the figure.  These same systems, however, have quite
small neutron star masses: the central values for PSR J0437-4715
\citep{vanstraten} and B1855+09 \citep{splaver2004} are only
1.55--1.60\msun, with upper limits at $\sim$1.7--1.75\msun. In
addition, the mass of PSR J1713+0747 (Splaver et al. 2005), at a
period of 68d (close to our evolutions 5) is only
1.3$\pm0.2$\msun. This value is three $\sigma$\ smaller than our
predicted mass of $\sim 2.01$\msun\ from sequence 5con, while
sequence 5eta provides a much more comfortable value of 1.655\msun,
as we have assumed the loss of half of the mass lost from the donor,
and, in addition, the initial mass of the NS is taken to be only
1.25\msun, contrary to the commonplace assumption of a starting value
of 1.35-1.40\msun\ often adopted in the literature. Notice that the
recent accurate mass determination of the relativistic double pulsar
system PSR J0737-3039 (Lyne et al. 2004) provides 1.25\msun\ for the
lighter component. To maintain small masses for the pulsar component,
we could also have reduced the companion mass down to 1\msun. When, on
the contrary the initial mass of the donor is somewhat larger than
1.1\msun, the first phases will occur at mass transfer rates much
larger than Eddington limit, and the additional initial mass will
easily be lost from the system. Of course, there is no adequate
physical reason to assume the semi--conservative evolution we adopt,
but a careful exploration of the parameter space is beyond the purpose
of the present work and the results are not affected sensibly by the
precise choices.  This is shown in Fig \ref{fig2}, which compares the
mass loss rate versus period evolution in cases 5con and 5eta.  In
this standard cases, after the super-- Eddington inital phase of mass
loss, a stationary phase begins, interrupted by the bump--induced
detachment, after which mass loss resumes and follows the same
rules. The difference between the conservative and non conservative
evolution is not very important, due to the fact that we assume that
the specific AM loss associated with the mass loss is that of the
primary neutron star, much smaller than the average AM for most of the
evolution.  The non conservative case, however, allows us to obtain in
the end small neutron star masses (1.6--1.7\msun) much more compatible
with the neutron star masses in the systems we are examining. The
conservative evolution leads to masses 1.9--2.1\msun.  Our following
discussion depends on the orbital evolution and not on the assumption
of mass conservation, as long as we assume AM is lost from the
neutron star.

\section{Evolution following the bump detachment: the role of radio--ejection}
\begin{figure}
\includegraphics[width=12cm]{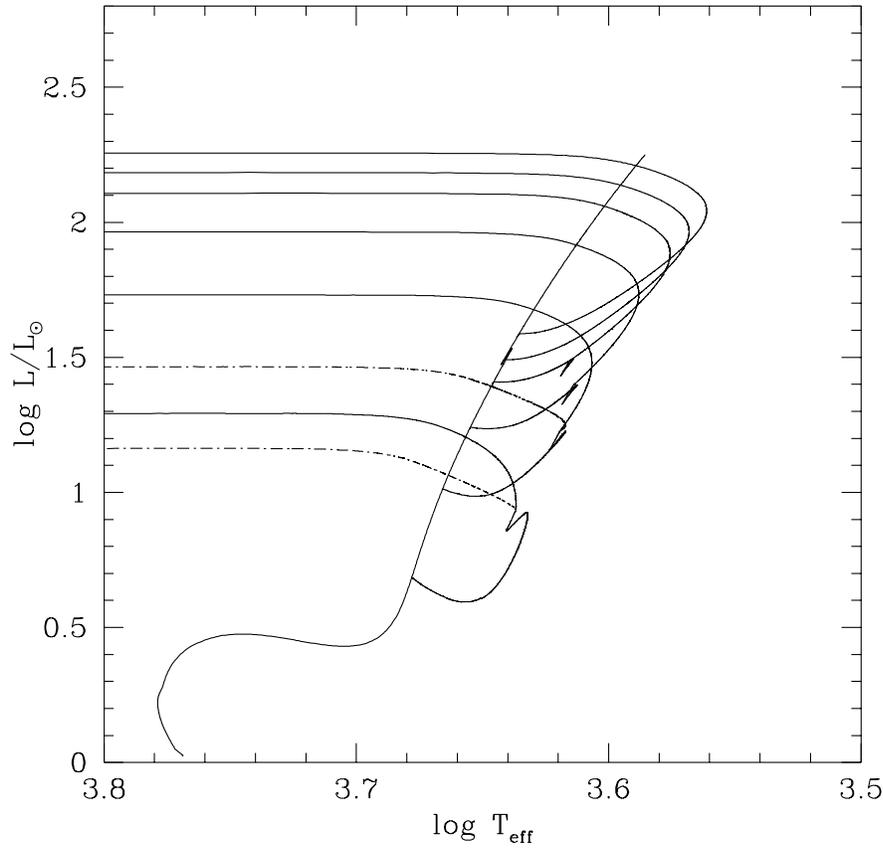} 
\figcaption[f3.eps]{HR diagram of
the single 1.1\msun\ evolution and of the binary evolutions from 3con
to 9con in Table 1, starting in order at increasing luminosities along the
track. The tracks starting below the bump show a bump signature. The
dashed curves starting at the bump of sequences 3con and 4con 
are sequences 3re and 4re08.
\label{fig3}}
\end{figure}

Fig. \ref{fig3} shows the HR diagram of the donor star 1.1\msun for
Z=0.02. The evolutionary track without mass loss is shown as a
reference. The location of the bump is seen in the thickening of the
track, at $\log L/L_\odot \sim 1.5$. It corresponds to a stage at
which the core mass is $\sim 0.218$\msun. We see that a bump signature
is present in all the evolutions where mass loss begins at
evolutionary phases at which the hydrogen burning shell has not yet
reached the discontinuity in hydrogen left by convection.

\begin{figure}
\includegraphics[width=10cm]{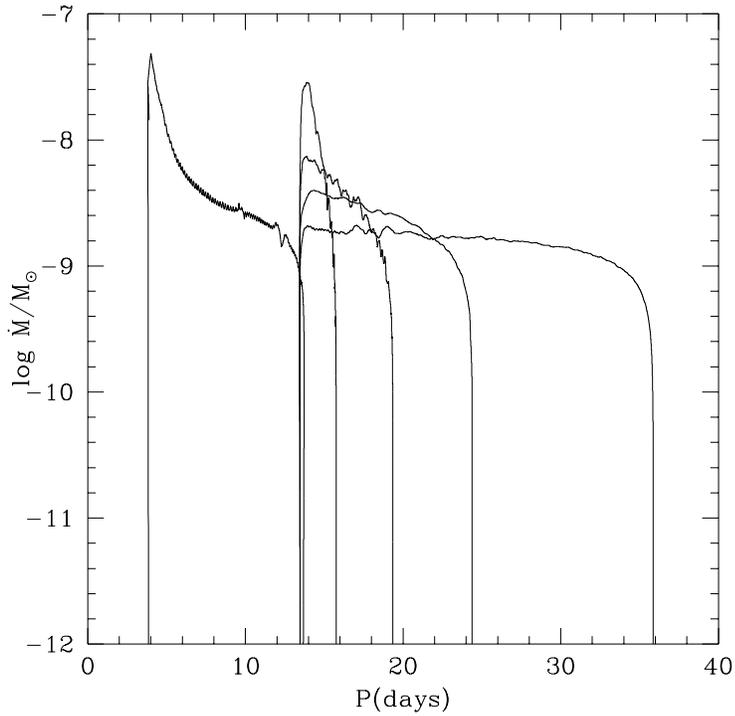} 
\figcaption[f4.eps]{ Mass
loss rate vs. orbital period for the ``standard" sequence 4, track
starting at the shortest period and ending, after the detachment, at the
longest period. After the bump-induced detachment,
three other tracks are shown, in which all the 
mass is lost from the system, carrying away 60\%, 80\% of
100\% of the specific AM of the donor.  The larger is the AM loss, the
larger is the mass transfer rate, and the shorter is the final orbital
period. If the AM loss is between 60 and 80\%, the final period is in the
range 20--24d.
\label{fig4}}
\end{figure}

At the bump--related detachment, in all cases, the NS has been
accelerated by the first phase of mass transfer. We assume that the
maximum spin achieved is the observed minimum spin period $P=1.3$ms.
During the semi detached evolution, the total mass decreases, while
the hydrogen shell advances and the H-exhausted core increases in
mass, until it will encounter the discontinuity. Of course the earlier
is the phase at which mass loss starts, the smaller is the total mass
at the time of detachment.  When the mass loss rate is resumed, it is
much larger if we assume that, thanks to an efficient radio--ejection,
the matter lost from the system carries away a large fraction of the
specific AM of the donor.  Figure \ref{fig4} shows the different
evolutions obtained in cases 4re.  The larger is the fraction of
specific AM lost, the larger is the mass loss rate, and the shorter is
the final \Porb.  We see that the standard evolution would lead to a
final period of $\sim$36 days, in the middle of the period gap. But if
radio ejection is efficient the period is reduced below 22 days for a
fraction between 60\% and 80\%$j_2$ of AM loss, corresponding to a
mass transfer rate of $\sim 5 \times 10^{-9}$\msun/yr.  We must then
understand when and where the mass transfer rates obtained are
compatible with the assumption of radio--ejection.

\begin{figure}
\includegraphics[width=12cm]{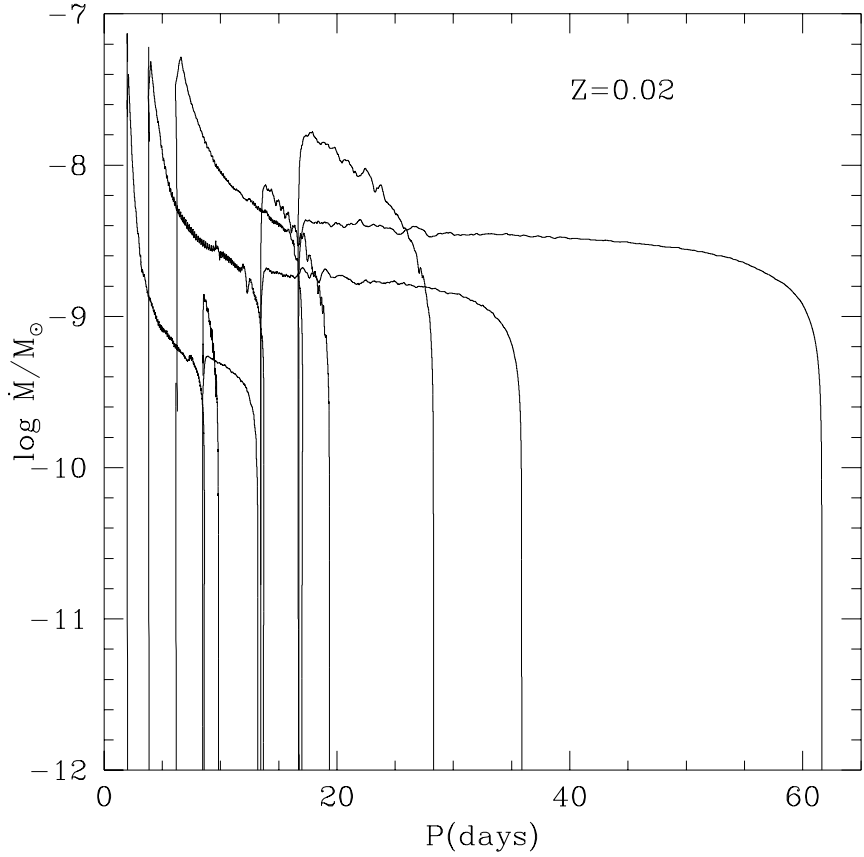} 
\figcaption[f5.eps]{ Mass loss
rate vs. orbital period for the cases 3con, 4con and 5con (from left to right). 
Starting at the bump detachment, in addition to the standard evolution, we show 
also the three sequences 3re, 4re08 and 5re08, in
which, due to radio-ejection, 80\% of the specific AM of the donor star is lost
with the mass lost by the system, and a shorter final orbital period is achieved.
\label{fig5}}
\end{figure}

Figure \ref{fig5} displays the mass transfer rates for cases 3, 4
and 5, and compares the standard evolution with the evolutions
assuming that 80\% of the donor AM is lost starting after the
bump--related detachment.  Contrary to the case discussed above, the
evolution of case 5 leads to a final period in the middle of the gap
($\sim$28d) if there is an efficient radio-ejection, corresponding to
a mass loss rate of 2$\times 10^{-8}$\msun/yr, otherwise the final
period is $\sim$61d, at the upper end of the period gap.
\\ 
\section{Constraints for the magnetic momentum of the pulsar}
\begin{figure}
\includegraphics[width=12cm]{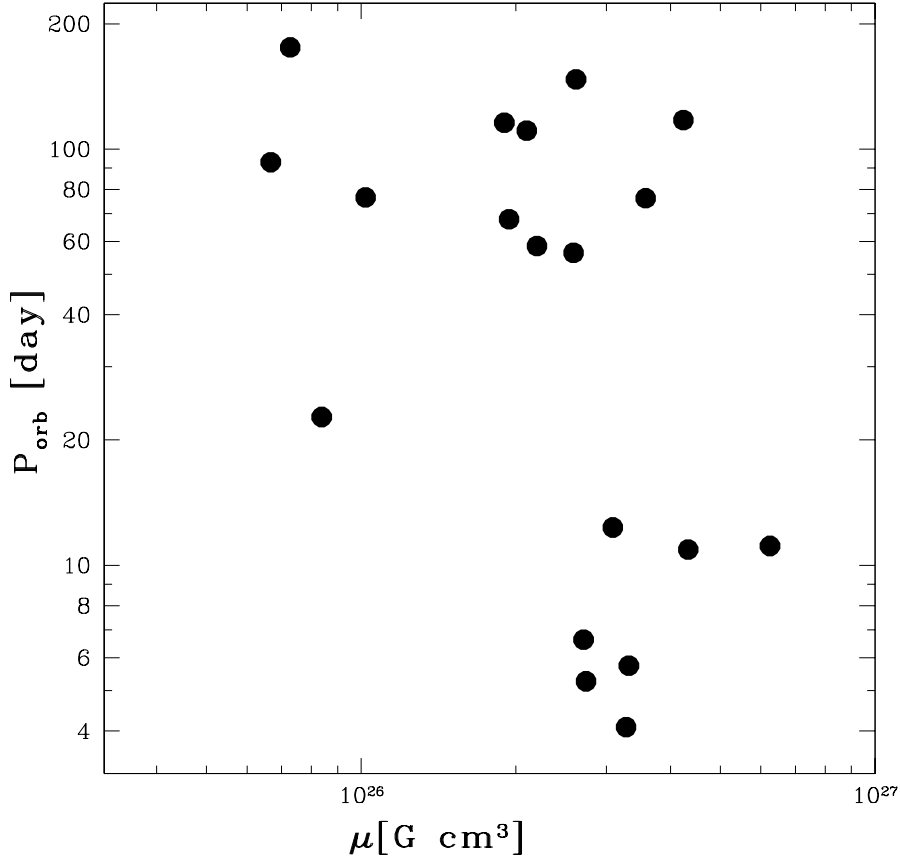} 
\figcaption[PvsMU.eps]{
Logarithmic plot of orbital periods $P_{\rm orb}$ vs. magnetic moments
$\mu$ for 19 galactic field MSPs in wide binary systems, whose
companion is expected to be a moderately massive white dwarf. The
other selection criteria are listed in the text. The shortage of
systems with orbital period in the interval between $\sim 20$ d and
$\sim 60$ d is evident. All the values of $\mu$ are calculated
assuming a moment of inertia $10^{45}$ g cm$^2$ for the neutron star
and using the formula $\mu=3.2\times 10^{37}\sqrt{P\dot{P}}$ G cm$^3,$
where $P$ and $\dot{P}$ are the pulsar spin period and its time
derivative.  Corrections to the value of $\dot{P}$ due to the pulsar
proper motion have been applied whenever they were available (12 cases
over 19).
\label{figdata}}
\end{figure}

Which are the constraints on the pulsar magnetic field which allow an
efficient radio--ejection in case 4 and non efficient in case 5?
We assume that the maximum spin frequency attainable by the MSP is
$\sim$750Hz \citep{chak2003} (corresponding to a minimum period of
$\sim$1.3ms). Following Burderi et al. (2001), we remember that
radio--ejection can occur for orbital periods longer than the
critical period:
\\ \\
\begin{eqnarray}
P_{\rm orb, crit} =  0.7 \times (\alpha^{-36} n_{0.615}^{-40})^{3/50} 
\dot{m}_{-10}^{51/25} m^{107/50} \times \nonumber \\
\mu_{26}^{-24/5} P_{spin, -3}^{48/5}\left[ 1 - 0.462 \left( {m_2} \over {m+m_2} 
\right)^{1/3} \right]^{-3/2}  \times \nonumber \\  (m+m_2)^{-1/2}
\;\; {\rm h}  &  & 
\label{eq:pcrit} 
\end{eqnarray}
where $\alpha$ is the Shakura--Sunyaev viscosity parameter,
$\mu_{26}$\ is the magnetic moment of the neutron star in units of
10$^{26}$Gauss cm$^3$ (corresponding to a surface magnetic field of
$10^8$Gauss), P$_{spin, -3}$\ is the spin period of the neutron star in
milliseconds, $n_{0.615} = n/0.615 \sim 1$ for a gas with solar
abundances (where $n$ is the mean particle mass in units of the proton
mass $m_{\rm p}$), $\dot{m}_{-10}$ is the accretion rate in units of
$10^{-10}$\msun/yr.  
This critical period has been derived equating the pressure of a
Shakura-Sunyaev accretion disc (approximately proportional to $r^{-2.6}$)
with the radiation pressure exerted by the magneto-dipole rotator
(proportional to $r^{-2}$). This gives a radius beyond which radiation
pressure always dominates over disk pressure, causing the ejection of
matter of the disc. Equating this radius with the distance between the NS 
and the inner lagrangian point, which we approximate with the difference between 
the orbital separation and the Roche lobe radius of the secondary
(which depends on the binary period), and solving for the orbital
period gives a critical period beyond which accretion is always inhibited
by the action of the pulsar pressure.
We can invert Equation 1 to obtain the value of the critical magnetic momentum 
as a function of the critical orbital period: we obtain
\begin{equation}
\mu_{26} = 0.93 A^{1/80} \dot m_{-10}^{51/120} m^{107/240} P_{spin,-3}^2
P_{orb,crit}^{-5/24} B^{-5/16} (m + m_2)^{-5/48} {\rm G cm^3}
\label{eq:two}
\end{equation}
where $A = \alpha^{-36} n_{0.615}^{-40}$, a value close to 1, and 
$B = 1 - 0.462 (m_2/(m+m_2))$.
Assuming that the NSs have been accelerated to the minimum observed
spin period $P_{spin,-3}=1.3$ms, inserting the appropriate values for the masses
(see table 1), and assuming that the orbital period at the bump is the
critical period, we can derive a constraint on the magnetic moment of
the NS. Namely, if we wish that the sequence 4 experiences
radio--ejection (to reduce its final period below 22d) while sequence
5 does not (to avoid a final period into the gap), we must impose:

\begin{equation}
2 \times 10^{26} \simlt \mu \simlt 4 \times 10^{26} {\rm G cm^3}
\label{eq:three}
\end{equation}
for those systems whose initial binary parameters would lead them into the
period gap. In this interpretation, the gap is only a statistical gap, and not
a forbidden region. Therefore, we can predict that systems with \Porb\ into
the gap can exist, but they should preferentially have magnetic moments
out of the interval given in Eq.\ref{eq:two}.
Fig.~\ref{figdata} reports the orbital periods vs. the magnetic
moments for all the known binary MSPs in the galactic field (as
derived from version 1.32 of the ATNF pulsar catalog\footnote{\tt
http://www.atnf.csiro.au/research/pulsar/psrcat}, and from Stairs et
al. 2005) having spin period shorter than 10 ms, $P_{\rm orb}>3$ d,
nearly circular orbit (eccentricity smaller than 10$^{-3}$) and
supposedly white dwarf companion of mass $\lsim 0.4$ M$_\sun.$ The
aforementioned parameters are the typical outcomes of the standard
nuclear evolution of low mass X-ray binary, which is in the focus of
this paper \citep[e.g.][]{webbink1983}.  It appears that about 75\% of
the sample displays magnetic moments within the range of equation
\ref{eq:two} or very close to its limits. The evolution of the 
three systems at \Porb$\sim$13 days might have been influenced by radio
ejection, and the two systems at \Porb$\sim$56 days could have been found
in the gap, if their magnetic moments had been larger.

\begin{figure}
\includegraphics[width=12cm]{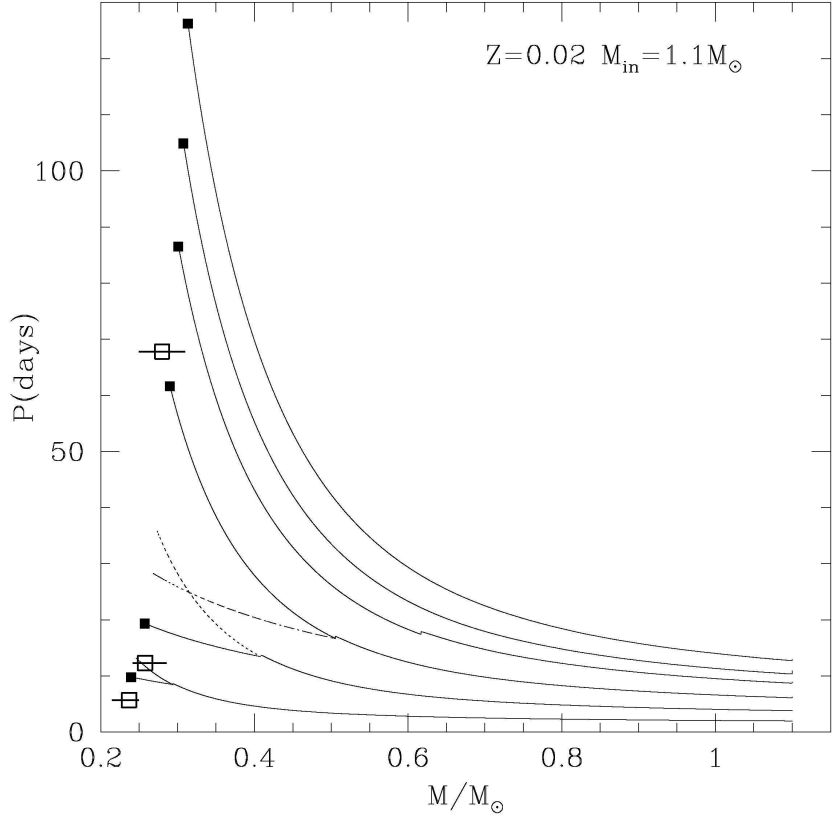} 
\figcaption[f6.eps]{ Orbital
period versus donor mass evolution.  For the first three tracks
starting from bottom, both the conservative case (3con, 4con and 5con, 
in order of increasing period)  
and the cases including radio--ejection, 3re, 4re08, 5re08, 
beginning at the bump related detachment and ending at a shorter period 
than the standard tracks) are shown.
At larger periods, the cases 6con , 8con and 9con are shown.  
In order to produce a period gap, as a consequence of the bump--related 
detachment, in the range 20 -- 60 days, the
radio--ejection phase should occur for the first and second system
from bottom, while it must not occur for the third system
(see text). Dots at the end of the sequences indicate which are the
white dwarf masses and orbital periods we expect to be result of the
evolutions. The three open squares are the systems
described in Fig. \ref{fig1}.
\label{fig6}}
\end{figure}

The proposed scenario is finally summarized in Fig. \ref{fig6},
reporting the evolution in the plane \Porb\ vs. white dwarf mass.  We
see that, taking into account the NS evolution and the process of
radio--ejection leads to a natural explanation of the period gap.  A
population synthesis may give further support to this picture,
allowing to precisely calculate the effect of the interplay between
the various parameters in eq. \ref{eq:pcrit} on the \Porb\ vs $\mu$
distribution.

\section{Population II models}
The result we have found derives from three main occurrences: 1) the
maximum extension of the convective envelope in population I reaches
down to an inner mass point of $\sim$0.24\msun, so that the bump
induced detachment occurs when the hydrogen exhausted core mass
reaches about 0.22\msun, at a period well below 20 days for the
systems in which we have tested that the radio--ejection mechanism can
work.  In population II, the maximum extent of convection only reaches
$\sim$0.35\msun. Thus any bump related detachment would occur at a
much longer longer orbital period, when the core reaches $\simgt
0.3$\msun\, and the red giant has already transfered a large amount of
its envelope. Therefore, even if radio--ejection begins to dominate,
it can not affect in an interesting way the binary final period. In
any case, no period gap can occur due to this mechanism, for periods
shorter than $\sim$35 days.

Unfortunately, the globular cluster period distribution
\citep{camilo-rasio} does indicate that it is affected by the
dynamic encounters in the dense stellar environment. In particular,
systems with period larger than 10 days are so few that we can not
falsify our hypothesis by now.

\section{Conclusions}
We follow the evolution of the mass losing component in systems
progenitors of binaries containing a MSP and a remnant white dwarf
companion. When the mass transfer begins during the red giant branch
phase, in many cases the binary evolution shows a period of
detachment, due to the shrinking of the stellar radius which occurs
when the hydrogen burning shell meets with the hydrogen discontinuity
left at the time of the maximum extension of the convective
envelope. When the mass transfer resumes, it is possible that the
evolution suffers a radio--ejection phase which alters the final
orbital parameters, due to the loss of mass and AM from the system. We
show that the period gap between 22 and 56 days in the distribution of
binary MSPs is produced if the magnetic momentum of the neutron star
is in the range $\sim 2 - 4 \times 10^{26}$Gcm$^3$, typical of MSPs.  A
population synthesis study could strengthen the conclusion that this
period gap is "bump--related", but there are many parameters to be considered in 
the analysis. A better test of the model would be if new systems are found into
the period gap, and their magnetic moments are either below 
$\sim 2 \times 10^{26}$Gcm$^3$, or above $\sim 4 \times 10^{26}$Gcm$^3$.

\acknowledgements
{This research was supported by PRIN 2003 ``Pushing ahead the frontiers 
of pulsar research". }

\clearpage

\begin{deluxetable}{lcccccccccccl}
\footnotesize
\tablecaption{Models \label{tbl-1}}
\tablewidth{0pt}
\tablehead{
\colhead{Seq.} & \colhead{$\eta^{a}$}&  \colhead{j$^{a}$} & 
\colhead{$N_{in}$} & \colhead{$M_{c, in}^{b}$} & 
\colhead{$M_{2,i}$} & \colhead{$M_{1,i}$} &\colhead{$M_{b}^{c}$} & \colhead{$M_{2,f}$}   &
\colhead{$M_{1,f}$} & \colhead{$P_{b}$(d)$^{c}$} &
\colhead{$P_{f}$(d)}  & \colhead{$\log (\dot{M})^{d}$}
}
\startdata
2con & 1 & j$_1$ & 200 &  0.0511  & 1.1  & 1.35 &     & 0.194 & 2.256 &     & 2.101 & -10.00  \\
3con & 1 & j$_1$ & 300 &  0.1550  & 1.1  & 1.35 &0.293& 0.246 & 2.072 &8.52 & 13.20 & -9.27  \\
4con & 1 & j$_1$ & 400 &  0.1868  & 1.1  & 1.35 &0.408& 0.273 & 2.047 &13.60& 35.86 & -8.67  \\
5con & 1 & j$_1$ & 500 &  0.2049  & 1.1  & 1.35 &0.505& 0.290 & 2.012 &16.80& 61.63 & -8.38  \\
6con & 1 & j$_1$ & 600 &  0.2184  & 1.1  & 1.35 &0.617& 0.301 & 1.977 &17.67& 86.57 & -8.11  \\
8con & 1 & j$_1$ & 800 &  0.2383  & 1.1  & 1.35 && 0.303 & 1.959 && 105.0 & -8.05  \\
9con & 1 & j$_1$ & 900 &  0.2463  & 1.1  & 1.35 && 0.310 & 1.939 && 126.3 & -7.85  \\
\tableline
3eta  &  .5 & j$_1$ & 300  & 0.1552 & 1.1 & 1.20 && 0.244 & 1.628 &8.14& 12.10 & -9.27    \\
4eta &  .5 & j$_1$ & 400 &  0.1868  & 1.1 &1.35 & 0.415 && 1.692 & 13.64 && -8.80  \\
5eta &  .5 & j$_1$ & 500 &  0.2049  & 1.1 &1.25 && 0.290 & 1.655 &16.84& 61.49 & -8.38  \\
\tableline
3re & 0 & 0.8j$_2$& 300  & 0.2106  & 0.293 & 2.025 && 0.240 & 2.025 && 9.817 & -8.84  \\
4re06 & 0 & 0.6j$_2$& 400  & 0.2266  & 0.408 & 1.692  && 0.263 & 1.692 && 24.38 & -8.42  \\
4re08 & 0 & 0.8j$_2$& 400  & 0.2266  & 0.408 & 1.912  && 0.257 & 1.912 && 19.35 & -8.08  \\
4re1 & 0  & j$_2$   & 400  & 0.2266  & 0.408 & 1.692  && 0.253 & 1.692 && 15.75 & -7.56  \\
5re08 & 0 & 0.8j$_2$& 500  & 0.2324  & 0.505 & 1.796  && 0.268 & 1.796 && 28.28 & -7.77  \\
5re1 & 0 & j$_2$   & 500  & 0.2367  & 0.505 & 1.547  && 0.261 & 1.547 && 21.46 & -7.00 \\
 \enddata
 
\tablenotetext{a}{$\eta=1$: conservative mass transfer up to
$\dot{M}_{Eddington}$, the mass exceeding the Eddington limit is lost
from the system.  $\eta=.5$: half of the mass lost by the donor is
accreted.  In both cases, the mass lost from the system takes away the
specific AM of the NS j$_1$. All the cases $\eta=0$ correspond to the
radio-ejection: all the mass lost from the donor is lost from the
system, carrying away a percentage of the donor specific AM, j$_2$,
specified in column 3.}  
\tablenotetext{b}{Helium core mass, in units
of \msun, when mass transfer begins.}  
\tablenotetext{c}{Mass of the donor and period at the bump-related detachment.}
\tablenotetext{d}{Units: M$_\odot$/yr. For the
sequences from 2 to 9, mass transfer rate in the stationary phase following
the peak at the beginning of mass transfer; for the radio--ejection sequences,
labelled "re", the peak mass
transfer rate at the beginning of the evolution is given.}
\end{deluxetable}

\end{document}